\documentclass{article}

\usepackage{amssymb}
\usepackage{amsmath}

\ifx\dvioutput\undefined
\usepackage[dvips]{graphicx}    % Êàðòèíêè
\else
\usepackage[pdftex]{graphicx}
\fi

\AtBeginDocument{
  \pretolerance-1
  \tolerance1900
  \adjdemerits10000
  \emergencystretch10mm   % Êíóò, ñ.131
  \clubpenalty10000       % øòðàô çà îòðûâ ïåðâîé ñòðîêè
  \widowpenalty10000
  \displaywidowpenalty10000
  }

\begin{document}

%%% remove comment delimiter ('%') and select language if required
%\selectlanguage{spanish}

\noindent \textbf{Microwave Photon Antibunching at the Modulation of the Resonance Frequency of a Qubit Emitter}

~\\~\\

A. P. Saiko${}^{1}$, R. Fedaruk${}^{2}$ and S. A. Markevich${}^{1}$
~\\

${}^{1}$Scientific-Practical Materials Research Centre NAS of Belarus, Minsk, Belarus

${}^{2}$\textit{${}^{ }$}Institute of Physics, University of Szczecin, 70-451, Szczecin, Poland

E-mail: saiko@ifttp.bas-net.by; fedaruk@wmf.univ.szczecin.pl
~\\~\\

\noindent PACS number(s): 42\textit{.}50\textit{.}Hz, 78.67.Hc, 81\textit{.}07\textit{.}Ta, 78.47.jp

~\\~\\
\begin{center}
Abstract
\end{center}

The statistics photons in the resonance fluorescence of a qubit excited by microwave and radio-frequency (RF) fields have been studied. It has been established that the coherent dissipative dynamics of the qubit with allowance for multiphoton emission and absorption of RF photons in each act of emission and absorption of a microwave photon results in periodic alternation of photon bunching and antibunching. It has been shown that periodically varying statistics photons can be smoothly transformed to the purely sub-Poisson statistics by varying the parameters of the RF field. The conditions for the formation of the «collapse-revival» of oscillations in the second-order correlation function photons have been determined. The described effects can be implemented on spin and superconducting qubits, quantum dots, and qubit-mechanical hybrid systems.
~\\~\\

The measurement of time correlation functions of radiation from light sources in quantum optics under lies the determination of its quantum properties and has a long history, including investigation of single photon emitters. One of the manifestations of non classical properties of light is photon antibunching. This phenomenon was predicted in [1--3] and was observed in resonance fluorescence from a monochromatic fieldexcited single atom in a beam [4, 5], a single ion in a trap [6], a single molecule in a crystal lattice [7], quantum dots [8], and nitrogenvacancy centers in diamond [9]. The characteristic feature of a signal of quantum correlations at quite intense excitation is its modulation by the Rabi frequency. Photon antibunching in fluorescence of light scattered by atoms excited by intense bichromatic light with nearly resonance frequencies was also studied [10].

Quantum properties of an emitter of single microwave photons, including photon antibunching in measurements of the secondorder correlation function for intensities, were recently observed with the use of artificial single atoms such as superconducting transmons [11, 12]. New features and capabilities of microwave quantum photonics can be manifested in studies of «dressed» states of qubits at their special bichromatic excitation. In this case, the microwave field transfers a qubit from the ground to excited state and the resonance frequency of the qubit is modulated by the radiofrequency (RF) field with the frequency close to the Rabi frequency of the qubit in the microwave field [13] or much higher than this Rabi frequency [14]. At the same time, such studies are of interest because of the possible application of such an excitation scheme in quantuminformation technologies [15, 16], to emulate the properties of hybrid spin-mechanical systems [17], to measure weak RF fields [18], and in quantum amplifiers and attenuators [19].

In this work, we study the correlation of photons emitted by spin qubits at their bichromatic excitation by the transverse microwave and longitudinal RF fields.

We consider a twolevel quantum system (e.g., spin qubit) in a microwave field inducing transitions between its states and in an RF field modulating its resonance frequency. The Hamiltonian of this system can be represented in the form [13, 14]

\begin{equation} \label{1}
H=H_{0} +H_{\bot } (t)+H_{\parallel } (t),
\end{equation}
 where $H_{0} =\omega _{0} s^{z} $ is the Hamiltonian of the qubit with the resonance frequency $\omega _{0} $; $H_{\bot } \left(t\right)=\omega _{1} (s^{+} +s^{-} )\cos \omega _{mw} t$ and $H_{\parallel } (t)=2\omega _{2} s^{z} \cos (\omega _{rf} t+\psi )$ are the Hamiltonians of the interaction of the qubit with the linearly polarized microwave and RF fields, respectively; $\omega _{1} $ and $\omega _{2} $ are the coupling constants; $\omega _{mw} $ and $\omega _{rf} $ are the frequencies of the microwave and RF fields, respectively; $\psi $ is the phase of the RF field; and $s^{\pm } $ and $s^{z} $ are the components of the spin (pseudospin) operator, which describe the ground (${\left| g \right\rangle} $) and excited (${\left| e \right\rangle} $) states of the qubit and satisfy the commutation relations $[s^{+} ,s^{-} ]=2s^{z} $, $[s^{z} ,s^{\pm } ]=\pm s^{\pm } $.

The dynamics of the qubit is described by the master equation for the density matrix $\rho $:
\begin{equation} \label{2}
i\hbar \partial \rho /\partial t=\left[H,\rho \right]+i\Lambda \rho
\end{equation}
(below, we set $\hbar =1$). Since usually $\omega _{1} /\omega _{mw} \ll 1$, the interaction of the qubit with the microwave field is considered in the rotating wave approximation [20]. The superoperator $\Lambda $describing decay processes is given by the expression [21] $\Lambda \rho =(\gamma _{21} /2)D[s^{-} ]\rho +(\gamma _{12} /2)D[s^{+} ]\rho +(\eta /2)D[s^{z} ]\rho $, where $\gamma _{21} $ and $\gamma _{12} $ are the rates of transitions from the excited state ${\left| e \right\rangle} $ of the qubit to its ground state ${\left| g \right\rangle} $ and vice versa, respectively; $\eta $ is the pure dephasing rate; $D[O]\rho =2O\rho O^{+} -O^{+} O\rho -\rho O^{+} O$ and $O$ is the spin operator.

We consider two different regimes of the dissipative dynamics of qubits.

(i) Let $\omega _{0} -\omega _{mw} -r\omega _{rf} \approx 0$ ($r=0,\pm 1,\pm 2,...$) and $\omega _{1} /\omega _{rf} $ be a small parameter. Then, the evolution of the quantum system with time-dependent Hamiltonian \eqref{1}, which is described by master equation \eqref{2}, can be considered within nonsecular perturbation theory with the use of Krylov--Bogoliubov--Mitropol'skii averaging [14] over fast oscillations after the application of the canonical transformation $u=\exp \left\{-i\left[\omega _{0} t+(2\omega _{2} /\omega _{rf} )\sin (\omega _{rf} t+\psi )\right]\right\}$ to Eq. \eqref{2}. After the implementation of both procedures, the equation for the modified density matrix $\tilde{\rho }_{r} $which describes the $r$th resonance, can be obtained from master equation \eqref{2} in the form $i\partial \tilde{\rho }_{r} /\partial t=\left[H_{eff} (r),\tilde{\rho }_{r} \right]+i\Lambda \tilde{\rho }_{r} $. Here, the time-independent effective Hamiltonian $H_{eff} (r)$ describes r-quantum transitions. As a result,
\begin{equation} \label{3}
H\to H_{eff} (r)=H_{0} (r)+H^{(1)} (r)+H^{(2)} (r),
\end{equation}
where
\[H_{0} (r)=(\omega _{0} -\omega _{mw} -r\omega _{rf} )s^{z} ,\]

\[H^{(1)} (r)=\frac{1}{2} \Omega (r)\left(s^{+} +s^{-} \right),\]

\[H^{(2)} (r)=\Delta _{BS} (r)s^{z} ,\]

\[\Omega (r)=\omega _{1} J_{-r} (z),\]

\[\Delta _{BS} (r)=\frac{1}{2} \sum _{n\ne r} \frac{\omega _{1}^{2} }{(r-n)\omega _{rf} } J_{n}^{2} (z),\]
$\Omega (r)=\omega _{1} J_{-r} (z)$ is the rf-field-renormalized Rabi frequency of the qubit in the microwave field, $\Delta _{BS} (r)$ is the Bloch--Siegert shift of the frequency of the resonance transition of the qubit (which is nonzero only for $r\ne 0$), and $z=2\omega _{2} /\omega _{rf} $ is the argument of the Bessel functions (which determines the magnitude of the effective interaction $\Omega (r)$ with the bichromatic field). Taking into account the Bloch--Siegert shift, we redefine the resonance conditions as $\omega _{0} +\Delta _{BS} (r)-\omega _{mw} -r\omega _{rf} \approx 0$. The solution of the master equation for the density matrix of the qubit $\tilde{\rho }_{r} $ in the reference frame rotating at the frequency $\omega _{mw} $ can be obtained in the explicit form (it is assumed that the qubit at the initial time $t=0$ was in the ground state ${\left| g \right\rangle} $):
\begin{equation} \label{4}
\tilde{\rho }_{r}^{rot} (t)=\frac{1}{2} -\frac{\gamma _{21} -\gamma _{12} }{2(\gamma ^{2} +\Omega _{\alpha }^{2} (r))} \times \left\{2(\gamma +\alpha )s^{z} +i\Omega (r)\left[s^{+} f_{r} (t)-H.c.\right]\right\}+
\end{equation}

\[+\left(\frac{e^{-i\Omega _{\alpha } (r)t} e^{-\gamma t} }{4\Omega _{\alpha } (r)} \left[\frac{i(\gamma _{21} -\gamma _{12} )}{\Omega _{\alpha } (r)-i\gamma } +1\right]\times \left[\left\{\Omega (r)\left[s^{+} f_{r} (t)-H.c.\right]-2\left[\Omega _{\alpha } (r)+i\alpha \right]s^{z} \right\}+H.c.\right]\right),\]

\noindent where $f_{r} (t)=\exp \left\{-i\left[r\psi +z\sin (\omega _{rf} t+\psi )\right]\right\}$, $\Omega _{\alpha } (r)=\sqrt{\Omega ^{2} (r)-\alpha ^{2} } $, $\alpha =(\gamma _{\bot } -\gamma _{\parallel } )/2$, $\gamma =(\gamma _{\bot } +\gamma _{\parallel } )/2$ and $\gamma _{\parallel } =\gamma _{12} +\gamma _{21} $ and $\gamma _{\bot } =(\gamma _{\parallel } +\eta )/2$ are the rates of energy and phase relaxations of the qubit, respectively (below, we set $\gamma _{12} \approx 0$).

From Eq. \eqref{4}, we obtain the following expression for the population of the excited level of the qubit absorbing one microwave photon with simultaneous absorption or emission of $r$ RF photons:
\begin{equation} \label{5}
{\left\langle e \right|} \tilde{\rho }_{r}^{rot} (t){\left| e \right\rangle} =\frac{1}{2} \frac{\Omega ^{2} (r)}{\Omega ^{2} (r)+\gamma _{\parallel } \gamma _{\bot } } \times
\end{equation}

\[\times \left\{1-e^{-\gamma t} \left[\cos \Omega _{\alpha } (r)t+\frac{\gamma _{\parallel } +\gamma _{\bot } }{2\Omega _{\alpha } (r)} \sin \Omega _{\alpha } (r)t\right]\right\}.\]
With the use of Eq. \eqref{4} and quantum regression theorem [20], an expression can be obtained for correlation functions of field amplitudes proportional to the spin operators of the qubit. The normalized second-order correlation function for the $r$th resonance can be represented in the form
\begin{equation} \label{6}
g_{r}^{(2)} (\tau )={\left\langle e \right|} \tilde{\rho }_{r}^{rot} (t){\left| e \right\rangle} \left|_{\rho (o)={\left| g \right\rangle} {\left\langle g \right|} } \right. /{\left\langle e \right|} \tilde{\rho }_{r}^{rot} (t\to \infty ){\left| e \right\rangle} ,
\end{equation}
where ${\left\langle e \right|} \tilde{\rho }_{r}^{rot} (t){\left| e \right\rangle} \left|_{\rho (o)={\left| g \right\rangle} {\left\langle g \right|} } \right. $ means that the solution of the master equation for the density matrix should be sought for the situation where the qubit at the initial time is in the ground state ${\left| g \right\rangle} $.

\noindent According to Eqs. \eqref{5} and \eqref{6}, oscillations of the correlation function $g_{r}^{(2)} (\tau )$ are caused by coherent oscillations of the population of excited levels of the emitter. For $\tau =0$, correlation is absent; i.e., $g_{r}^{(2)} (0)=0$. This indicates the quantum nature of radiation leading to the photon antibunching effect. In the case under consideration of the excitation of the qubit by the bichromatic field, emission and reemission are complicated. The modulating RF field modifies the energy spectrum of the system: the ground and excited states of the qubit are supplemented by levels splittings between which are specified by the frequency $\omega _{rf} $ (Fig. 1a) [14]. Owing to the modified spectrum of the qubit at the main resonance ($\omega _{0} -\omega _{mw} =0$), multiphoton absorption of a certain number of RF photons and

\begin{center}
\includegraphics*[]{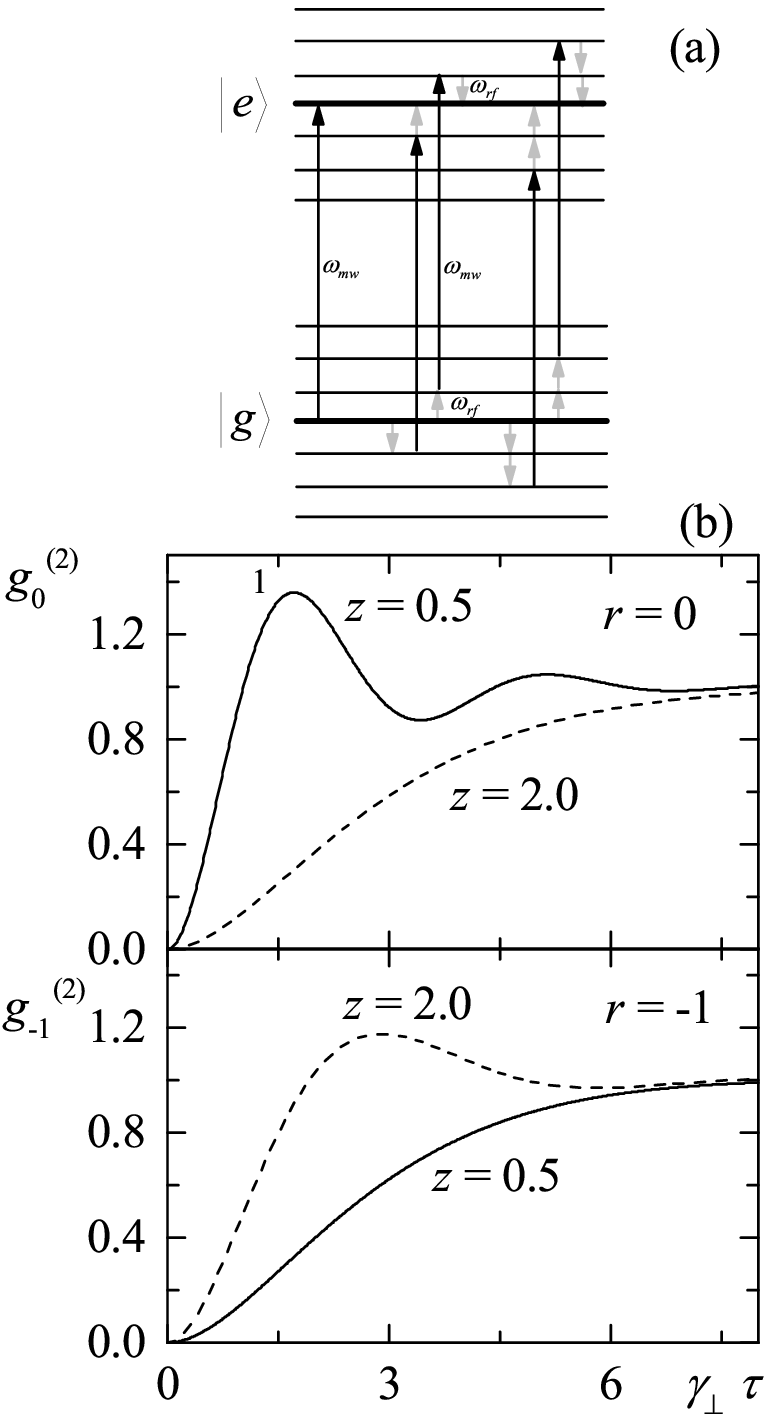}
\end{center}

Fig. 1. (a) Scheme of multiphoton transitions excited by bichromatic radiation in the case $r{\rm \; }={\rm \; }0$. Thick lines correspond to the levels of the initial spin system that is not modified by radiation. (b) $g_{r}^{(2)} (\tau )$ for $r=0,-1$ at bichromatic excitation of qubit, $\omega _{mw} =\omega _{0} $, $\omega _{1} /2\pi =0.1$ MHz, $\omega _{rf} /2\pi =1$ MHz, $z=0.5$ è $z=2.0$, $\gamma _{\bot } =0.05$ MHz, $\gamma _{\parallel } =0.01$ MHz. \\

\noindent emission of the same number of photons occur simultaneously with the emission or absorption of a microwave photon. These processes are taken into account in the zeroth order Bessel function and reduce the probability of absorption of a microwave photon owing to the renormalization of the Rabi frequency: $\omega _{1} \to \Omega (0)=\omega _{1} J_{0} (z)$, $z=2\omega _{2} /\omega _{rf} $. With an increase in the parameter $z$, the function $J_{0} (z)$ decreases and is zero at $z_{1} \approx 2.41$, $z_{2} \approx 5.52$, etc. Consequently, with an increase in $z$, the effective frequency of oscillations $\Omega _{\alpha } (0)=\sqrt{\omega _{1}^{2} J_{0}^{2} (z)-\alpha ^{2} } $ decreases (if $\omega _{1} J_{0} (0)>\alpha $), becomes zero at a certain $z$ value when $\omega _{1} J_{0} (z)=\alpha $, and then becomes imaginary and the oscillation--relaxation behavior is replaced by the relaxation behavior (see Figs. 1b and 2a). Therefore, purely sub-Poisson statistics are established instead of periodically varying statistics of photons in a time window comparable to the relaxation time $1/\gamma $.

\begin{center}
\includegraphics*[]{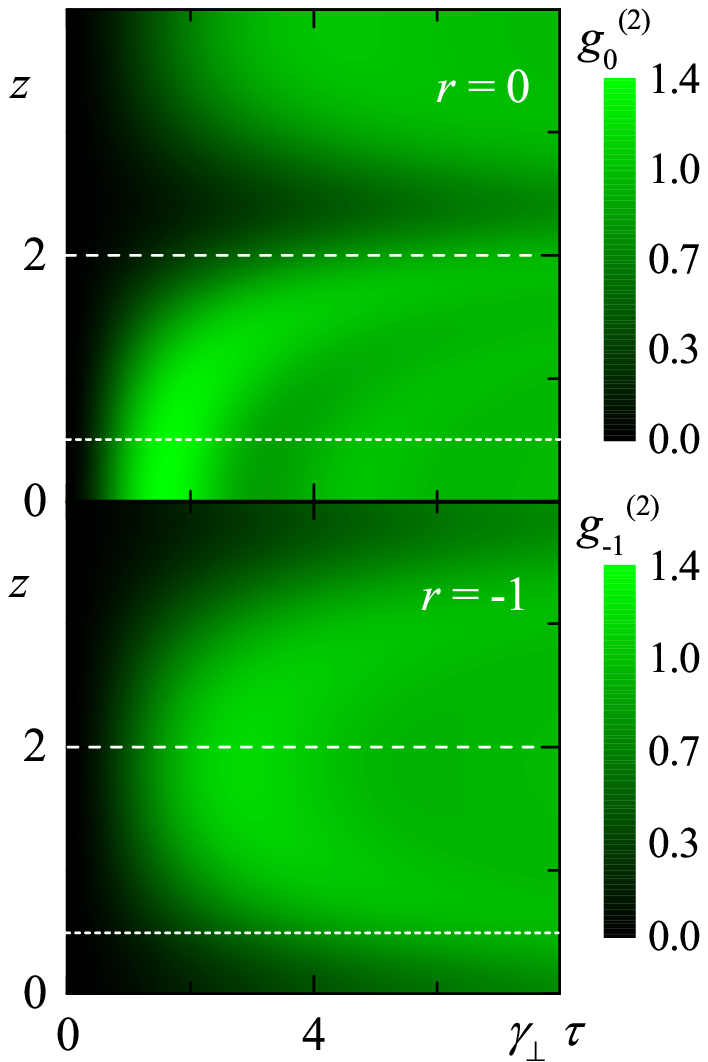}
\end{center}

Fig. 2. (Color online) Correlation function $g_{r}^{(2)} (\tau )$ at bichromatic excitation of the qubit versus$z=2\omega _{2} /\omega _{rf} $ for $\omega _{0} +\Delta _{BS} (r)-\omega _{mw} -r\omega _{rf} \approx 0$, $\omega _{1} /2\pi =0.1$ MHz, $\gamma _{\bot } =0.05$ MHz, $\gamma _{\parallel } =0.01$ MHz. (a) $r=0$. (b) $r=-1$. The dotted ($z=0.5$) and dashed ($z=2.0$) lines indicate the positions of the signals shown in Fig. 1b. \\

We note that the role of the function $J_{0} (z)$, which appears owing to multiphoton absorption and emission of the same number of RF photons, is similar to the role of the Debye--Waller factor in solid state physics. This factor, e.g., takes into account the elastic scattering of phonons at the interaction of the crystal lattice with electrons, which leads to the formation of phononless lines in optical spectra. In this respect, oscillations of the nanomechanical system interacting with the qubit can serve as the RF field; as a result, the properties of, e.g., the hybrid spin--mechanical system can be emulated by means of the modulating RF field.

\noindent At side resonances, $\omega _{0} +\Delta _{BS} (r)-\omega _{mw} \mp \left|r\right|\omega _{rf} =0$, the absorption of a microwave photon is accompanied by the absorption (--) or emission (+) of RF photons whose number is larger than the number of emitted or absorbed RF photons by $\left|r\right|$. In this case, the frequency of oscillations of the function $g_{r}^{(2)} (\tau )$ is given by the expression $\Omega _{\alpha } (r)=\sqrt{\Omega ^{2} (r)-\alpha ^{2} } $. The implementation of both the main and side resonances is accompanied by multiphoton absorption and emission of the same number of RF photons, which is taken into account in the Bessel functions in the expressions for renormalized Rabi frequencies, $\Omega (r)=\omega _{1} J_{-r} (z)$.

\noindent Variation of these frequencies owing to the variation of the parameters of the RF field ($z$ value) is manifested in the oscillation behavior of the correlation function $g_{r}^{(2)} (\tau )$, which vanishes in the limit $\tau \to 0$ (Figs. 1b and 2b). As is seen, in contrast to the main resonance, an increase in z for the side resonance results in the transition from the sub-Poisson statistics of photons to periodically varying statistics. Thus, in contrast to the monochromatic excitation, photons are emitted and absorbed cascade-by-cascade rather than one-by-one. As a result, the experimentally observed emission and further absorption of only one microwave photon at the main resonance and for the case of side resonances, e.g., for $r=-1$, the emission and absorption of one microwave photon and one RF photon occur with smaller probabilities because of the corresponding factors $J_{0}^{2} (z)$ and $J_{-1}^{2} (z)$,taking into account directly unobserved processes of multiphoton absorption and emission of RF photons.

(ii) We assume that $\omega _{1} >>\omega _{2} $, $\omega _{2} /\omega _{rf} $is a small parameter, the microwave field is strong ($\omega _{1}^{2} \gg \gamma _{\parallel } \gamma _{\bot } $), and the conditions of the exact resonance of the microwave field with the qubit ($\omega _{mw} =\omega _{0} $) and Rabi resonance ($\omega _{1} =\omega _{rf} $) are satisfied. Then, the density matrix of the qubit in the bichromatic field can be represented in the rotating reference frame in the form

\noindent
\begin{equation} \label{7}
\rho _{rot} (t)=\frac{1}{2} -\frac{1}{4} \cos \psi e^{-\gamma t} 
\end{equation}

\[\times \left[e^{i(\omega _{1} t+\psi )} (s^{+} -s^{-} +2s^{z} )-H.c.\right]+\frac{i}{8} \sin \psi  \]

\[\times \left(R(t)\left\{\left[e^{i\left(\omega _{1} t+\psi \right)} (s^{+} -s^{-} +2s^{z} )-H.c.\right]+2(s^{+} +s^{-} )\right\}\right. \]

\[\left. + R*(t)\left\{\left[e^{i\left(\omega _{1} t+\psi \right)} (s^{+} -s^{-} +2s^{z} )-H.c.\right]-2(s^{+} +s^{-} )\right\}\right)\]

\noindent where $R(t)=e^{-\gamma 't} \left(\cos \Omega _{\beta } t-\frac{\beta +i\omega _{2} }{\Omega _{\beta } } \sin \Omega _{\beta } t\right)$, $\Omega _{\beta } =\sqrt{\omega _{2}^{2} -\beta ^{2} } ,$ $\gamma =(\gamma _{\bot } +\gamma _{\parallel } )/2,$ $\gamma '=\gamma -\beta $ and $\beta =\gamma _{\parallel } /4.$ Using the regression theorem [20] and Eq. \eqref{7}, we obtain the normalized second-order correlation function
\begin{equation} \label{8}
g^{(2)} (\tau )=1-e^{-\gamma \tau } \cos \psi \cos (\omega _{1} \tau +\psi )
\end{equation}

\[-\frac{1}{2} \sin \psi e^{-\gamma 't} \left(\sin \left[(\omega _{1} +\Omega _{\beta } )\tau +\psi \right]^{^{} } +\sin \left[(\omega _{1} -\Omega _{\beta } )\tau +\psi \right]\right. \]

\[\left. +\frac{\beta }{\Omega _{\beta } } \left[\cos \left[(\omega _{1} +\Omega _{\beta } )\tau +\psi \right]-\cos \left[(\omega _{1} -\Omega _{\beta } )\tau +\psi \right]\right]\right).\]
According to Eq. \eqref{8}, $g_{0}^{(2)} (0)=0$; i.e., the radiation of the qubit is of a quantum character. The microwave field at the bichromatic excitation induces Rabi oscillations with the frequency $\omega _{1} $ and splits each of the levels of the qubit into two sublevels with the energy gap $\omega _{1} $, whereas the RF field induces Rabi oscillations with the frequency $\omega _{2} $ and splits each of the formed sublevels into two sublevels with the energy gap $\omega _{2} $. Figure 3a illustrates the double splitting of the upper level of the qubit in the bichromatic field. As a result, the emission spectrum of the (qubit + bichromatic field) system includes three Mollow triplets: $\omega _{mw} $, $\omega _{mw} \pm \omega _{2} $; $\omega _{mw} +\omega _{1} $, $\omega _{mw} +\omega _{1} \pm \omega _{2} $ and $\omega _{mw} -\omega _{1} $, $\omega _{mw} -\omega _{1} \pm \omega _{2} $ [16]. Oscillatory terms in Eq. \eqref{8} for the correlation function $g^{(2)} (\tau )$ correspond to quantum transitions with the frequencies $\omega _{1} $ and $\omega _{1} \pm \omega _{2} $. As is shown in Fig. 3b, the collapse--revival effect is clearly manifested in oscillations of $g^{(2)} (\tau )$ ; i.e., the intensity of radiation of the qubit vanishes periodically at the times $\tau _{n} =(2n+1)\pi /\Omega _{\beta } $ and then is restored. This occurs because oscillations of the population difference between the qubit levels at the frequency $\omega _{1} $ are modulated by rf-field-induced Rabi oscillations with the frequency $\Omega _{\beta } $. Change in the amplitude of the RF field can retard or hasten the development of collapse and revival of oscillations in the correlation function, thus changing the statistics of emitted photons (Fig. 4). The oscillatory behavior of $g^{(2)} (\tau )$, including the collapse--revival effect, is sensitive to the phase of the RF field (see Eq. (8) and Fig. 3b). At $\psi =0$, the effect of the RF field in-phase with the microwave field is completely eliminated and oscillations occur only at the frequency $\omega _{1} $. If the phase of the RF field is shifted by $\pi /2$ with respect to the phase of the microwave field, oscillations at the frequency $\omega _{1} $ disappear and oscillations appear at the sum ($\omega _{1} +\Omega _{\beta } $) and difference ($\omega _{1} -\Omega _{\beta } $) frequencies, leading to the collapse--revival effect. At a random phase of the RF field, Eq. \eqref{8} is averaged over the uniform distribution of $\psi $ from $0$ to $2\pi $. In this case, oscillations at three frequencies ($\omega _{1} $, $\omega _{1} \pm \Omega _{\beta } $) make comparable contributions and the collapse--revival effect is most pronounced. In the laboratory reference frame, groups of microwave and RF photons at the frequencies corresponding to three Mollow triplets are involved in emission and absorption. If the RF field is absent, Eq. \eqref{8} is reduced to the known expression [3, 20] for the case of monochromatic excitation when the correlation function oscillates with the frequency $\omega _{1} $.

Thus, the second-order correlation function for resonance fluorescence of the qubit excited by the microwave and RF fields has the properties confirming the emitted photon antibunching effect. It was established that emission and absorption acts are multiphoton. At $\omega _{1} \ll \omega _{rf} $, multiphoton emission and absorption of RF photons occur simultaneously with radiation or absorption of each microwave photon. These processes can be controlled by the RF field and by varying the behavior of the correlation function until the complete suppression of oscillations and transition to the antibunching regime determined by the dephasing rate. At $\omega _{1} =\omega _{rf} $, there is the effect of collapse--revival of oscillations of the correlation function caused by R-field-induced Rabi oscillations between dressed states of the qubit. The predicted properties of the correlation function can be potentially applied not only for spin qubits but also for artificial single atoms, such as superconducting qubits or quantum dots, as well as for qubit-mechanical hybrid systems.

\noindent

\begin{center}
\includegraphics*[]{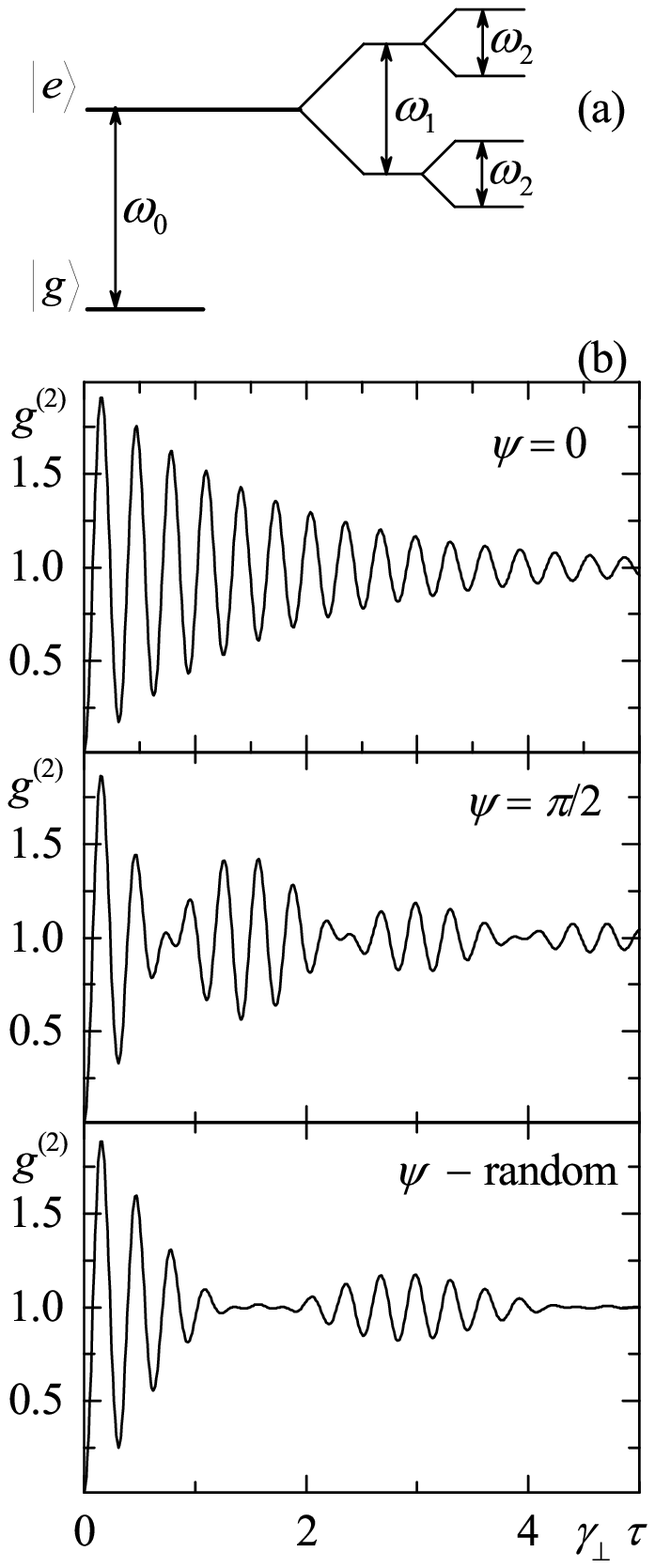}
\end{center}
 
Fig. 3. (a) Scheme of transitions excited by bichromatic radiation in the case $\omega _{rf} \approx \omega _{1} $ in the rotating reference frame. (b) Correlation function $g^{(2)} (\tau )$ at $\omega _{mw} =\omega _{0} $, $\omega _{rf} =\omega _{1} $, $\omega _{1} /2\pi $ = 1 MHz, $\omega _{2} /2\pi $= 0.1 MHz, $\gamma _{\bot } =0.05$ MHz, $\gamma _{\parallel } =0.01$ MHz. \\

\begin{center}
\includegraphics*[]{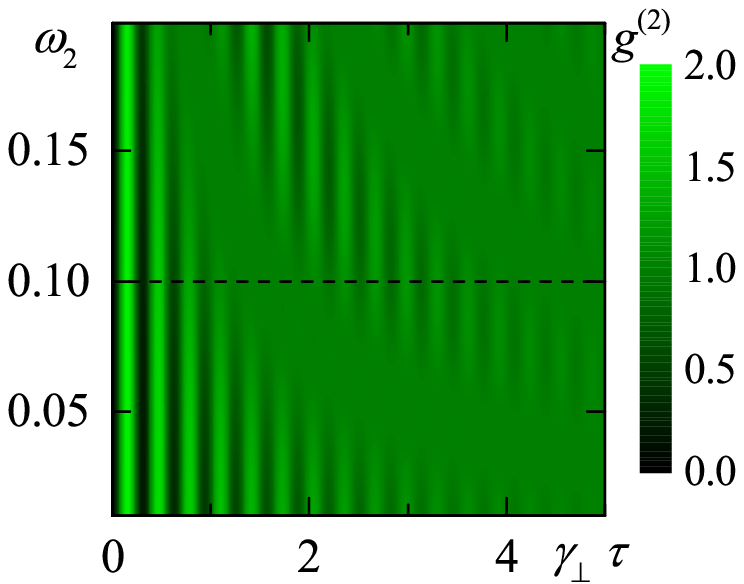}
\end{center}
 
Fig. 4. (Color online) Correlation function $g^{(2)} (\tau )$ versus $\omega _{2} $ at $\omega _{mw} =\omega _{0} $, $\omega _{rf} =\omega _{1} $\textbf{,} $\omega _{1} /2\pi $ = 1 MHz, $\gamma _{\bot } =0.05$ MHz, $\gamma _{\parallel } =0.01$ MHz, $\psi $ is a random variable. The dashed line corresponds to the signal presented in Fig. 3b. \\

\noindent \textbf{References}

\begin{enumerate}
\item  H. J. Carmichael and D. F. Walls, J. Phys. B \textbf{9}, 1199 (1976).

\item  H. J. Kimble and L. Mandel, Phys. Rev. A \textbf{13}, 2133 (1976).

\item  C. Cohen-Tannoudji and S. Reynaud, Phil. Trans. R. Soc.\textit{ }A \textbf{293}, 223 (1979).

\item  H. J. Kimble, M. Dagenais and L.Mandel, Phys. Rev. Lett. \textbf{39}, 691 (1977).

\item  A. Aspect, G. Roger, S. Reynaud, J. Dalibard and C. Cohen-Tannoudji, Phys. Rev. Lett.\textit{ }\textbf{45}, 617 (1980).

\item  F. Diedrich and H. Walther, Phys. Rev. Lett. \textbf{58}, 203 (1987).

\item  T. Basche, W. E. Moerner, M.Orrit, and H. Talon, Phys. Rev. Lett. \textbf{69}, 1516 (1992).

\item  P. Michler, A. Kiraz, C. Becher, W. V. Schoenfeld, P. M. Petroff, Lidong Zhang, E. Hu, and A. Imamoglu, Science \textbf{290}, 2282 (2000).

\item  C. Kurtsiefer, S. Mayer, P. Zarda, and H. Weinfurter, Phys. Rev. Lett. \textbf{85}, 290 (2000).

\item  Y. Ben-Aryeh, H. Freedhoff and T. Rudolph, J. Opt. B \textbf{1}, 624 (1999).

\item  D. Bozyigit, C. Lang, L. Steffen, J. M. Fink, C. Eichler, M. Baur, R. Bianchetti, P. J. Leek, S. Filipp, M. P. da Silva, A. Blais and A.Wallraff, Nature Physics \textbf{7}, 154 (2011).

\item  I.-C. Hoi, T. Palomaki, J. Lindkvist, G. Johansson, P. Delsing and C. M. Wilson, Phys. Rev. Lett.\textit{ }\textbf{108, }263601 (2012).

\item  A. P. Saiko and G. G. Fedoruk, JETP Lett. \textbf{87}, 128 (2008).

\item  A. P. Saiko, G. G. Fedoruk, and S. A. Markevich, JETP\textbf{ 105}, 893 (2007).

\item  L. Childress and J. McIntyre, Phys. Rev. A \textbf{82}, 033839 (2010).

\item  A. P. Saiko and R. Fedaruk, JETP Lett.\textit{ }\textbf{91}, 681 (2010).

\item  S. Rohr, E. Dupont-Ferrier, B. Pigeau, P. Verlot, V. Jacques and O. Arcizet, Phys. Rev. Lett. \textbf{112} 010502 (2014).

\item  M. Loretz, T. Rosskopf and C. L. Degen, Phys. Rev. Lett.\textbf{ 110}, 017602 (2013).

\item  S. N. Shevchenko, G. Oelsner, Ya. S. Greenberg, P. Macha, D. S. Karpov, M. Grajcar, A. N. Omelyanchouk and E. Il'ichev, Phys. Rev. B \textbf{89}, 184504 (2014).

\item  M. O. Scully and M. S. Zubairy, \textit{Quantum Optics,} Cambridge University Press, Cambridge, 1997.

\item  A.P. Saiko, R. Fedaruk and S. A. Markevich, J. Phys. B \textbf{47}, 155502 (2014).
\end{enumerate}

\noindent

\end{document}